\documentclass[12pt,a4paper]{article}
\usepackage[T1]{fontenc}
\usepackage{epsfig}
\usepackage{apalike}
\usepackage{amssymb}
\usepackage[centertags,intlimits]{amsmath}
\author{Jean-Marc ALLAIN and Martine BEN~AMAR\\
\normalsize{Laboratoire de Physique Statistique, CNRS-UMR 8550,}
\\\normalsize{Ecole Normale Sup\'erieure, 24, rue Lhomond 75231 Paris France}}

\begin{document}

\title{Budding and Fission of a multiphase vesicle}
\maketitle

\begin{abstract}
We present a model of bi-phasic vesicle in the limit of large surface 
tension. In this regime, the vesicle is
completely stretched and  well described by two spherical caps with a 
fold  which concentrates the  membrane stress.
The conservation laws and geometric constraints restrict the space of 
possible shapes to a pair of
solutions labeled by a parameter $\tau$ given by  {\it line 
tension/pressure}. For a  given $\tau$ value, the two
solutions differ by the length of the interface between domains. For 
a critical value
$\tau_c$ , the two vesicle shapes become identical and no solution  exists
above this critical value. This model sheds new light on  two 
proposed mechanisms (osmotic shocks and molecule
absorption) to explain  the budding and the fission  in recent experiments.
\end{abstract}

\section{Introduction}

The cell membrane is a bilayer made out of a mixture of lipid 
species. The membrane is both the boundary of the cell
  and an interface inside the cell, separating different compartments. 
This soft structure is responsible for many
biological properties. Intracellular traffic is also realized by 
membrane structures: a membrane vesicle buds from one
compartment, travels through the cytosol and fuses with another 
compartment. Despite the fluidity of the lipid bilayer,
the cellular membrane presents a lateral inhomogeneity due the 
formation of dynamical microdomains, called rafts
\cite{Simons97}. These microdomains have been shown to be rich in 
cholesterol and sphingolipid \cite{Brown00}. {\it In
vivo}, the rafts have not been directly observed but their size has 
been estimated to be between 20 and 700 nm
\cite{Chazal03}. A central question in membrane biology and 
biophysics is to understand how this spatial organization is
used by the cell, in particular to favor interactions with proteins. 
Due to their size and specific composition, it has
been argued that rafts   play a role in protein docking, signaling, 
intracellular traffic \cite{VanMeer04} or virus
budding \cite{Chazal03}.

         Recently, a model system of Giant Unilammelar Vesicles (GUV) 
including sphingomyelin-cholesterol domains
  was developed \cite{Dietrich01}. These domains, which are supposed 
to reproduce raft composition, are the result of a
phase separation of the lipid species \cite{Veatch03}. They are more 
structured than the surrounding classical liquid
bilayer but remain in a liquid state. For this reason, they are 
called "liquid-ordered" domains whereas the classical
membrane is called "liquid-disordered". A large number of studies 
have focused on the thermodynamic of liquid-ordered
phases, in particular the effect of temperature or composition 
changes on domain formation \cite{deAlmeida03,Veatch03}.
Multi-phase vesicles are elegant and efficient tools to study the 
mechanical properties of microdomains. It can be used
to understand how rafts bud and make daughter vesicles for 
intracellular traffic, but also how detergent addition can
isolate rafts from the cell membrane. Recent experiments have shown 
that liquid-ordered domains can be separated from the
initial vesicle by using tubular deformations \cite{Allain04a}, 
osmotic shocks \cite{Baumgart03,Bassereau05} or
absorption of external molecules like proteins or detergents 
\cite{Staneva04,Staneva05}. Here, we develop a macroscopic
theory for the two last situations. Our model describes the budding 
preceeding the fission where the
liquid-ordered domain is lift up from the liquid-disordered vesicle.

         Budding and fission have already attracted many theoretical 
works for homogeneous
  \cite{Jaric95,Seifert97,Dobereiner97,Tanaka04,Sens04} or inhomogeneous
\cite{Seifert93,Julicher96,Kohyama03,Laradji04,Harden05} membranes. 
The  models vary depending on the physical
interactions involved but they are all based on the minimization of 
the bilayer energy \cite{Helfrich73}. Due to the
non-linearity of the steady-shape equations, a numerical treatment is 
often required. We focus our attention
on multi-phase vesicles slightly stretched, a situation often 
encountered in experiments. In this case, osmotic pressure
effects dominate and we show that the vesicle can be described by two 
spherical caps with an elastic junction. The
variational procedure with surface constraints allows to find two 
solutions for any ratio {\it $\tau=$ line
tension/pressure} less than a critical value
$\tau_c$.  The stable solution is the one observed experimentally. 
An osmotic shock increases the control parameter
$\tau$ and so destabilizes the stable solution which may lead the 
system to a complete fission of the neck. The case of
detergents is slightly different since it requires an energy model 
for molecular absorption in the membrane. When
detergent molecules are added in the membrane, they locally deform 
the bilayer. According to Leibler's model
\cite{Leibler86}, these curvature defects can be taken into account 
by a term in the energy proportional to both the
average curvature and the concentration of molecules. Homogeneous 
concentration of molecules is favored  away from the
interface between domains. At the junction, a concentration gradient 
appears. If the chemical inhomogeneity is localized
at the junction, the addition of molecules leads to an increase of 
the effective line tension inducing a budding and
  a possible separation into two independent vesicles.

          Our model explains qualitatively and even quantitatively the 
budding and  fission created
  by osmotic shocks or proteins absorption. It is a physical approach 
based on continuum description and its domain of
validity ends at the molecular level. Because of its simplicity, 
extension and application to other processes may be
achieved easily.

\section{Membrane description.}

\subsection{Inhomogeneous lipid bilayer.}

         We consider an inhomogeneous vesicle constituted by two lipid 
phases: a 'liquid-ordered' ($l_o$)
and a 'liquid-disordered' ($l_d$). Both phases are in the liquid 
state but the ($l_o$) domain is more structured than the
($l_d$) phase due to the following reasons: there are specific 
interactions between molecules \cite{Li01} and/or there is
an optimization of biphilic space packing \cite{Holopainen04}. Steady 
morphologies and their out-of-plane deformations
are well described by the Canham and Helfrich's model with energy for 
each phase $i$ given by:
\begin{equation}
F_m^i = \int_S {\left[ {2\kappa_i{\sf H}^2 + \kappa_G^{(i)}{\sf K} + 
\Sigma_i} \right] dS }
\label{eqn:Fm}
\end{equation}
${\sf H}$ and  ${\sf K}$ are respectively the mean and Gaussian 
curvature. The elastic bending rigidity
$\kappa_i$ and Gaussian rigidity $\kappa_G^{(i)}$ are expected to be 
higher in the $l_o$ phase. Typical values can be
found for example in \cite{HandbookLipids}: 
$\kappa_{l_d}\simeq20{\text k}_{\text b} {\text T}$ and
$\kappa_{l_o}\simeq80{\text k}_{\text b} {\text T}$. Values of 
Gaussian moduli are notoriously more difficult to measure
but a recent study mentions values of order $\kappa_G^{(i)}= - 
0.83\kappa_i$ \cite{Siegel04}. Although $F_m^i$ is a
surface integral, the Gaussian contribution to the energy is indeed a 
contour integral calculated at the interface
between the two domains, due to the Gauss-Bonnet theorem.

         The last contribution in Eq.\ref{eqn:Fm} is related to the 
possible extension of the membrane.
In the case of a stretched vesicle, this contribution is large 
compared to the elastic energy and the membrane surface
can be considered as constant. This is taken into account by 
introducing the Lagrange multiplier $\Sigma_i$.

         The total energy of the two-domain vesicle includes the 
energy (\ref{eqn:Fm}) of each phase plus
two coupling terms. First, a sharp interface of vanishing thickness 
exists between the $l_o$ and $l_d$ phases. Any
increase of its length requires an energy proportional to a line 
tension $\sigma$. Second, the vesicle membrane is
lightly permeable to the water but not to the ions or big molecules 
present in the surrounding water medium. This induces
an osmotic pressure $P$. The energy of the coupling terms is:
\begin{equation}
\label{eqn:Fcoup}
F_c = \sigma\int_C{dl} - P\int{dV}
\end{equation}

\subsection{Proteins or detergent-membrane interactions}

         External molecules such as proteins or detergents can be 
absorbed in both phases but with different
  efficiencies. Their introduction in the membrane is well described 
by a Landau's approach with an optimal homogeneous
concentration $\phi_{eq}$. Departure from this value or inhomogeneity 
of concentration $\phi$ has a cost in energy,
assumed quadratic to leading order. The energy cost is given by two 
positive constants in each phase: $\alpha_i$ and
$\beta_i$. If the proteins are soluble or not in the surrounding 
medium, we can either set the number of these molecules
in each phase or set the chemical potential $\mu_i$ of the phase $i$. 
We choose to fix $\mu_i$ but this has no real
incidence on the results since it only affects the definition of 
$\mu_i$. Therefore the free chemical energy of
absorption for each phase is:
\begin{equation}
\label{eqn:Fprot}
F_{p}^i = \int_{S_i}{ \left( {\frac{\alpha_i}{2}(\phi-\phi_{eq_i})^2 
+ \frac{\beta_i}{2}(\nabla \phi)^2 } + \mu_i\phi \right)  dS} + 
\int_{S_i}{ \Lambda_i {\sf H}\phi  dS}
\end{equation}
The last integral in Eq.\ref{eqn:Fprot} represents the local 
distortion of the membrane induced by
the absorbed molecules \cite{Leibler86,Bickel01}. It is proportional 
to the mean curvature of the membrane with a weight
depending on the local concentration $\phi$, as suggested by 
S.~Leibler \cite{Leibler86}, $\Lambda_i$ being a coupling
constant. The absorption process itself  affects differently the two 
leaflets of the vesicle. We restrict our attention
to the case where the adsorption takes place on one side only. In 
such case, $\Lambda_i$ is positive on the outer side
absorption and negative on the inner side. When the two layers are 
affected by absorption, two concentration fields are
necessary and our theoretical framework can be easily adapted to 
address such situation. Taken into  account all
previous contributions, the total free energy for the system is given by:
\begin{equation}
\label{eqn:FTOT}
F_{TOT} = F_m^o + F_m^d + F_p^o+ F_p^d + F_c
\end{equation}
The usual variation procedure to identify extrema of this energy 
produces the so called Euler-Lagrange equations.

\subsection{Euler-Lagrange equations.}

\begin{figure}[p]
\begin{center}
          \includegraphics[width=3.25in]{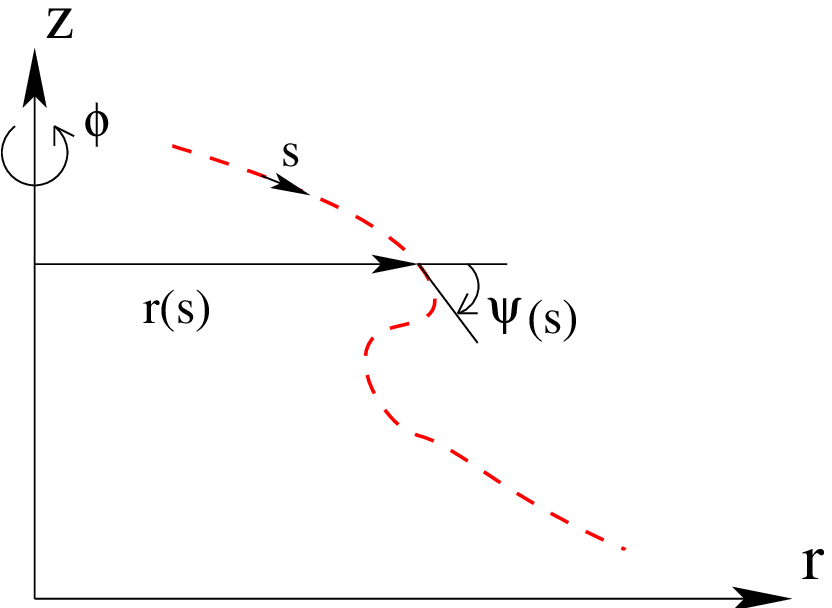}
          \caption{General parameterization of an axisymmetric vesicle
in cylindrical coordinates. The dashed curve
is the membrane. The parameterization is done by the arc-length $s$.
The shape of the membrane is given by $r(s)$
and $\psi(s)$. The two domains have the same
parameterization.\label{fig:param}}
\end{center}
\end{figure}

         Minimization of the free energy $F_{TOT}$ gives the static 
solutions for the membrane. Looking for
axisymmetric shapes, we choose the cylindrical coordinates and we 
parameterize the surface by the arc-length $s$. The
vesicle shape is given by $r(s)$ and $\psi(s)$ (see 
Fig.~\ref{fig:param}). We have derived the Euler-Lagrange equations
associated with $F_{TOT}$ \cite{Allain04b}. They are
\begin{subequations}
\label{eqn:ELg}
\begin{eqnarray}
         \psi '' &=& \frac{\sin(\psi)\cos(\psi)}{r^2} - \frac{\psi 
'}{r}\cos(\psi) - \frac{Pr}{2\kappa_i}\cos(\psi) +
\frac{\gamma}{\kappa _ir}\sin(\psi)  + \frac{\Lambda_i}{\kappa_i}\phi' \\
         \gamma ' &=& \frac{\kappa _i}{2}{\psi'}^2 - 
\frac{\kappa}{2r^2}\sin(\psi)^2 + \tilde{\Sigma}_i - Pr\sin(\psi) - 
\Lambda_i\phi\psi' + \frac{\alpha_i}{2}\phi^2  \\
&&+ \frac{\beta_i}{2}\phi'^2 + \tilde{\mu}_i\phi \nonumber \\
         \phi'' &=& - \phi'\frac{\cos(\psi)}{r} 
-\frac{\Lambda_i}{\beta_i}\left( {\frac{\sin(\psi)}{r} + 
\psi'}\right) + \frac{\alpha_i}{\beta_i}\phi 
+\frac{\tilde{\mu}_i}{\beta_i} \\
         r' &=& \cos(\psi).
\end{eqnarray}
\end{subequations}
These equations have to be solved with the suitable boundary 
conditions at the border between the two domains. To simplify the 
notations, we introduce the following parameters: $\tilde{\Sigma}_i = 
\Sigma_i + \alpha_i/2\phi_{eq_i}^2$ and $\tilde{\mu}_i = \mu_i - 
\alpha_i\phi_{eq_i}$. Assuming continuity of both the radius $r$, the 
angle $\psi$ and the molecules concentration $\phi$, the variational 
procedure gives also three equations for the boundary conditions:
\begin{subequations}
\label{eqn:CLg}
\begin{eqnarray}
         \lefteqn{\kappa _{1}\psi '(s_J-\epsilon)r(s_J) + (\kappa_{1}+ 
\kappa _{G_1})\sin(\psi(s_J)) - \Lambda_1\phi(s_J)r(s_J)}\\
         &&- \kappa _{2}\psi '(s_J+\epsilon)r(s_J) - 
(\kappa_{2}+\kappa _{G_2})\sin(\psi(s_J)) + \Lambda_2\phi(s_J)r(s_J) 
= 0, \nonumber\\
         \lefteqn{\gamma (s_J - \epsilon) - \gamma (s_J + \epsilon) + 
\sigma = 0} \\
\lefteqn{\beta_1\phi'(s_J-\epsilon)-\beta_2\phi'(s_J+\epsilon) = 0.}
\end{eqnarray}
\end{subequations}
where $s_J$ is the arc-length at the junction, label $1$ denotes the 
phase for $s\le s_J$ and label $2$
  the phase for $s\ge s_J$.

         Since these equations are highly non-linear, there is no 
exact solutions for arbitrary
values of the physical parameters. However further analytical 
progress can be obtained in the limit of large pressure
(stretched membrane). Remarkably, this treatment only requires simple 
analytical algebra and allows to explain
experimental features such as the complete budding of the ordered 
phase obtained by different groups using either osmotic
shocks \cite{Baumgart03}, proteins \cite{Staneva04} or detergent 
molecules \cite{Staneva05}

\section{Analytical treatment of the membrane shape.}

         We first consider a membrane without absorbed molecules. A 
solution of the Euler-Lagrange equations can
be easily found if we discard the contribution from the elasticity. 
We use this simple solution as zeroth order and
correct it for small but not vanishing values of the bending rigidity 
by using boundary layer analysis. We consider also
the inclusion of molecules with no chemical activity. They are 
described in the model by curvature defects. For a weak
coupling between curvature and concentration, a similar strategy is 
used to understand how the molecules affect the
membrane shape.

\subsection{The exact zero-order model: the  capillary solution.}

\begin{figure}[p]
\begin{center}
          \includegraphics[width=3.25in]{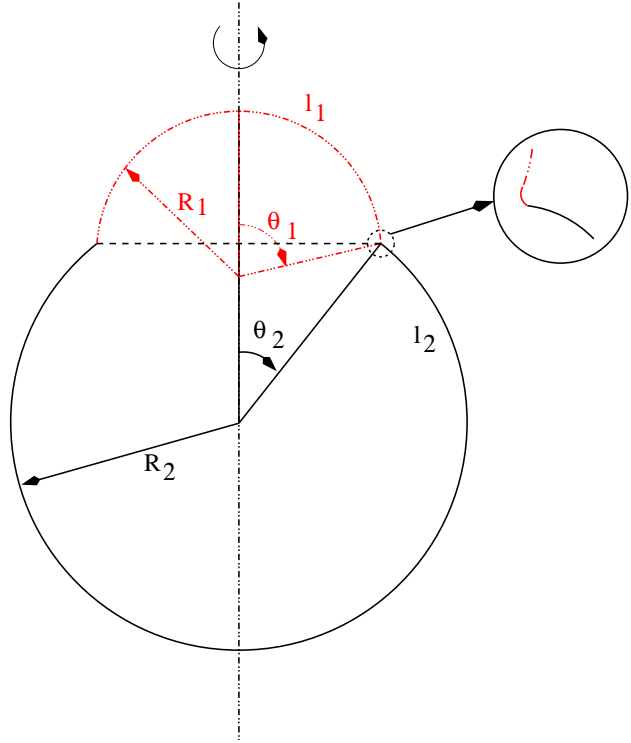}
          \caption{Schematic representation of a  axisymmetric 
vesicle, including the four parameters $R_1$,
$R_2$, $\theta_1$ and $\theta_2$ used in the vesicle description. The 
circle details the fold near the interface, where
the elastic properties can no longer be neglected.
          \label{fig:coord_fold}}
\end{center}
\end{figure}

            For stretched membrane without absorbed molecules ($\phi = 
0$), it is believed that after electro-formation of GUV vesicles the 
osmotic pressure dominates the elastic energy. When $\kappa_i= 0$ in 
both phases, a solution of the Euler-Lagrange equation is made of two 
spherical caps defined by a set of four geometrical parameters: the 
radii of the two caps $R_1$, $R_2$ and the two angles at the boundary 
$\theta_1$ and $\theta_2$ (see Fig.~\ref{fig:coord_fold}). The 
contact between the two caps gives a first continuity relation
\begin{equation}
\label{eqn:EL_tense_1}
R_1\sin\theta_1 = R_2\sin\theta_2.
\end{equation}
The Euler-Lagrange equations~(Eq.\ref{eqn:ELg}) give the values of 
the two Lagrange multipliers $\Sigma_i$ and $\gamma_i$, without 
direct information on the vesicle shape:
\begin{subequations}
\begin{eqnarray}
&&2\Sigma_i = PR_i, \\
&&\gamma_i(s) = \frac{PR_i^2}{2}\sin\psi\cos\psi.
\end{eqnarray}
\end{subequations}
the angle $\psi$ being proportional to the arc-length $s$. Only the 
boundary condition and the
conservation of the area of each phase give the possibility to fix 
completely the ideal shape. From Eq.\ref{eqn:CLg}, we
deduce:
\begin{equation}
\label{eqn:EL_tense_2}
R_1^2\sin\theta_1\cos\theta_1 = 
R_2^2\sin\theta_2\cos\theta_2-\frac{2\sigma}{P}.
\end{equation}
The shape is  controlled by the reduced line tension $\tau = 
\sigma/P$ (homogeneous to a surface),
which can be adjusted by changing the osmotic pressure. As an 
example, from the figure (1b) in Baumgart et al.'s work
\cite{Baumgart03}, reproduced here in figure~\ref{fig:Baumgart}, we 
calculate $\tau =20.5 \mu m^2$, from estimated values
of $R_1$, $R_2$, $\theta_1$ and $\theta_2$. Notice that in Baumgart's 
work, label $1$ correspond to the $l_d$ domain and
label $2$ to the $l_o$ domain.

\begin{figure}[p]
\begin{center}
          \includegraphics[width=2in]{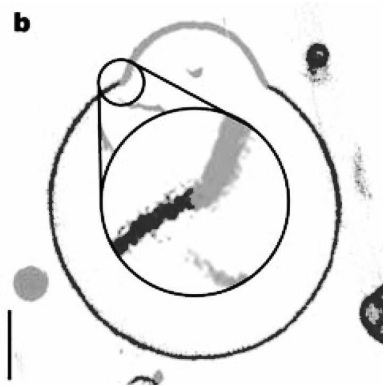}
          \caption{Black and white version of figure (1b) from 
Baumgart et al.'s work \cite{Baumgart03}.
The picture is a two-photon microscopy image, showing equatorial 
section of GUVs with two coexisting domains.
The $l_o$
domain appears in grey here and the $l_d$ in dark. Scale bar, $5\ \mu 
m$. Reproduced from \cite{Baumgart03} with the
authorizations of the authors and editor. \label{fig:sol1}}
\end{center}
\end{figure}

\begin{figure}[p]
\begin{center}
          \includegraphics[width=3.25in]{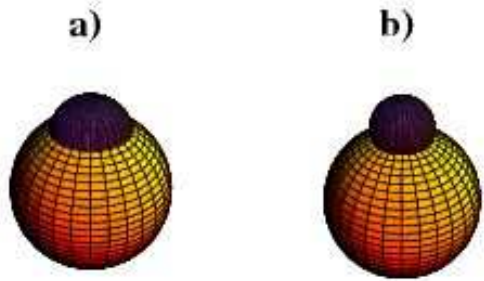}
          \caption{The two solutions of the E.-L. 
equations~(Eq.\ref{eqn:EL_tense_1} and \ref{eqn:EL_tense_2}) for
$\tau = 20.5\mu m^2$. The two pictures have the same scale. (a) 
Experimental solution \cite{Baumgart03}, with
$R_1=5.3 \mu m$, $R_2=11 \mu m$, $\theta_1 = 1.3$ and $\theta_2 = 
0.51$. The associated dimensionless
energy, given by Eq.\ref{eqn:FTOT_tense} is $\tilde{F}_{TOT}=-1.380$. 
(b) Calculated solution with $R_1= 4.0
\mu{\text m}$, $R_2=10 \mu{\text m}$, $\theta_1 = 2.0$ and $\theta_2 
= 0.36$. The dimensionless energy of vesicle (b) is
$\tilde{F}_{TOT}=-1.377$, meaning that the solution is experimentally unstable.
          \label{fig:Baumgart}}
\end{center}
\end{figure}

          Solving Eq.\ref{eqn:EL_tense_1} and \ref{eqn:EL_tense_2} for 
the above $\tau$-value gives two possible solutions: $R_1=5.30 \mu 
m$, $R_2=10.5 \mu m$, $\theta_1 = 1.34$ and $\theta_2 = 0.514$, the 
measured values (Fig.~\ref{fig:sol1}(a)) but also $R_1= 3.97 \mu m$, 
$R_2=10.3 \mu m$, $\theta_1 = 1.96$ and $\theta_2 = 0.364$ for the 
second solution (Fig.~\ref{fig:sol1}(b)). In order to explain why the 
first solution is preferred in the experiment, we calculate the 
energy which is restricted here to two contributions: $F_{TOT} = -PV 
+ \sigma l$ with $l$ the perimeter of the interface. Using a typical 
length scale $L_r = 10 \mu m$, it is possible to construct
the dimensionless energy $\tilde{F}_{TOT} = F_{TOT}/(\pi PL_r^3)$. 
Notice that the value of $L_r$ does not affect the
physics of the problem, it is used only to have dimensionless lengths 
close to $1$. So we obtain
\begin{equation}
\label{eqn:FTOT_tense}
\tilde{F}_{TOT} = - V/\pi L_r^3+ 2\tau R_1\sin\theta_1/L_r^3,
\end{equation}
which gives respectively (-1.380) compared to (-1.377), and shows 
that the experimental
observed solution is stable
while the other one is unstable as expected.

\begin{figure}[p]
\begin{center}
\includegraphics[width=3in]{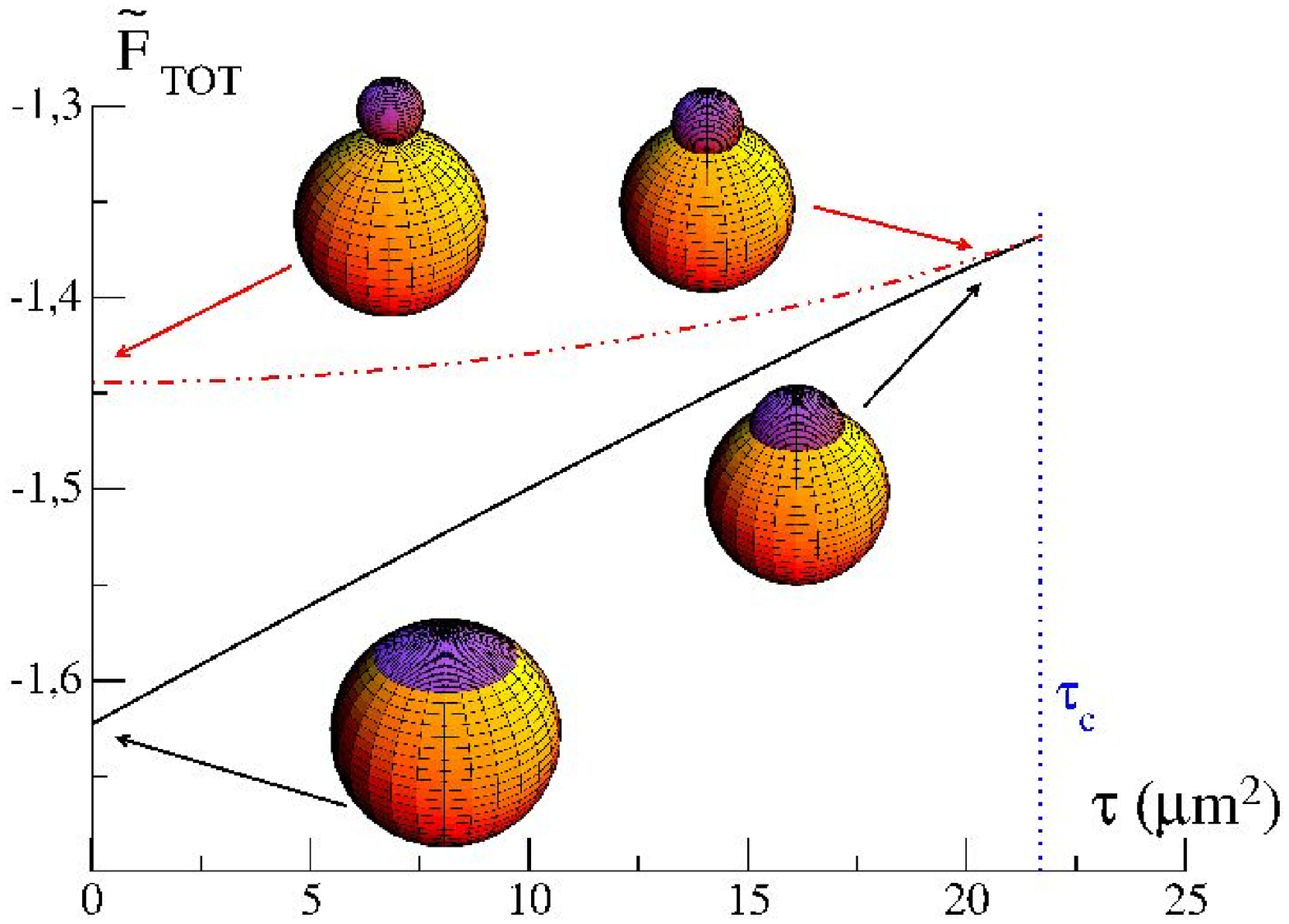}\\
(a)\\
\vspace{0.5cm}
\includegraphics[width=3in]{rJ_tau.eps}\\
(b)
          \caption{Dimensionless energies $\tilde{F}_{TOT}$ (figure a)
and interface radius $r_J$ (figure b) of the solutions of the 
E.L.~equations~(Eq.\ref{eqn:EL_tense_1}
and \ref{eqn:EL_tense_2}) versus the control parameter $\tau = 
\sigma/P$. The calculation has been done with the area
$A_1 = 136 \mu m^2$ and $A_2 = 1296 \mu m^2$. In both figures, the 
solid line corresponds to the stable solution,
experimentally observable, the dashed line to the unstable solution 
and the dotted line to $\tilde{\tau} =
\tilde{\tau}_c$, the critical value of the control parameter. For 
$\tilde{\tau} \ge \tilde{\tau}_c$, there is no longer a
solution. Four pictures of vesicles showing the shape transformation 
with $\tau$ have been added.
          \label{fig:ener_sol1}}
\end{center}
\end{figure}

         A systematic study of the pair of solutions for arbitrary 
values of $\tau$ is straightforward and the results
are presented in figure~\ref{fig:ener_sol1}. 
Figure~\ref{fig:ener_sol1}a is a classical bifurcation diagram when a 
pair
of solutions appears with opposite stability. In this problem, $\tau$ 
is the control parameter and the energy
$\tilde{F}_{TOT}$ is the order parameter. These two solutions differ 
geometrically, the unstable solution presenting a
smaller neck compared to the stable one (obvious if $\tau=0$) (see 
Fig~\ref{fig:ener_sol1}b). As $\tau$ increases, the
two solutions become geometrically closer up to a finite value of 
$\tau_c=R_1 R_2/2$. Above the critical value ($\tau \ge
\tau_c$), there is no connected solution of the Euler-Lagrange 
equations but the solution with two separated spheres
remains.

         This bifurcation diagram describing change in the topology of 
budding spheres is similar to the one found in
the catenoid problem where a soap film is fixed on two parallel 
rings, separated by a small distance $d$ compared to the
radius of the ring. Two different  minimal surfaces  (with similar 
catenoid shapes) satisfy the variational equations
derived from the capillary energy. The difference between these two 
shapes can be measured by the perimeter at
mid-distance between the two rings. The catenoid with the smaller 
neck is unstable since its area is larger and,
experimentally, the other catenoid is observed. However, as it is 
well known, as the distance $d$ increases, the neck
size decreases, the catenoid is destroyed and is replaced by two 
independent disks \cite{BenAmar98} with topology
changes. This geometrical instability is not reversible. At the 
fission, the neck of the catenoid is not zero but the
analytical calculation shows that the two catenoids, the stable and 
the unstable, have the same shape.

         In our case, we are faced with the same type of capillary 
instability where there exist two similar solutions
whose stabilities are governed by the energy. As the control 
parameter, here the effective line tension, is increased,
the two solutions merge and a change of topology is expected at this 
point. We do not know if this change is irreversible
since fission requires microscopic reorganization such as hemifission 
\cite{Kozlovsky03}. Experimentally, the daughter
vesicles can remain connected by a small filament of lipids but if 
the two vesicles move away, the process is of course
not reversible. The critical value $\tau_c$ is determined by the 
fourth equation in Eq.\ref{eqn:EL_tense_2} which gives
the equilibrium of the forces in the radial direction (axis $r$). The 
term in $\tau = \sigma/P$ is due to the line
tension and its effect is to pinch the membrane. The two others terms 
(in $R_1^2$ and $R_2^2$) are related to the
pressure force on the membrane and are bounded. The critical value 
$\tau_c$ is the value for the maximal force on the
membrane. For higher line tension (or smaller pressure), it is no 
longer possible to compensate for the line tension
which splits the system into two independent vesicles.

\begin{figure}[p]
\begin{center}
          \includegraphics[width=3.25in]{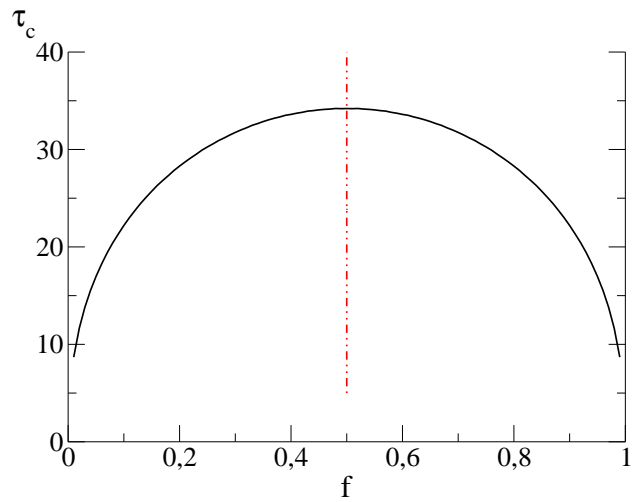}
          \caption{Values of the critical control parameter $\tau_c$ 
versus the fraction $f$ of the upper domain
(label $1$). The calculation has been done with a fixed total area 
$A_{tot} = A_1 + A_2 = 1433 \mu m^2$. The areas of the
domains are given by $A_1 = f\,A_{tot}$ and $A_2 = (1-f)A_{tot}$. The 
dashed line shows the fraction $0.5$.
          \label{fig:limit_sol1}}
\end{center}
\end{figure}

          One important conclusion of this study is the fact that 
small domains are more easily ejected. This can be
validated or invalidated experimentally when a vesicle has several 
$l_o$ domains of various size. This conclusion is
opposite to a floppy membrane whose shape is controlled by elasticity 
\cite{Lipowsky03}. To show this, we have varied the
fraction $f$ of the upper domain (label $1$) and we have calculated 
$\tau_c$ using the data of the experimental example
(see Fig.\ref{fig:limit_sol1}). Since the two phases are equivalent 
when elasticity is neglected, the results are the
same for $f$ and $1-f$. The parameter $\tau_c$ increases with the 
size of the smallest domain. This result can be
explained by a simple argument in the limiting case of a flat domain 
on a flat surface. If the radius of the domain is
$r$, the pinching energy (due to the line tension) is approximatively 
$\sigma r$ and the resistance energy (due to the
pressure) is approximatively $Pr^3$. The balance of the two energies 
gives $\sigma/P \approx r^2$. Therefore, it is
harder to destabilize a large domain than a small one. Next, we study 
the robustness of the model when elasticity is
taken into account.

\subsection{The elasticity localization.}

         Comparing the bending energy (Eq.\ref{eqn:Fm}) to the osmotic 
pressure energy (Eq.\ref{eqn:Fcoup}) one finds
that elastic effect can be neglected if $\kappa_i<<P R_i^3$ in each 
phase. However, a discontinuity of the tangent
appears at the interface between the two domains creating a 
singularity in the curvature. As soon as the bending modulus
is exactly zero, this discontinuity produces an infinite elastic 
energy contribution, localized near the junction, in
contradiction with the weakness of elasticity. We are faced with a 
classical boundary layer model, as found for example
in the crumpling of an elastic plate \cite{BenAmar97} or the folding 
of an elastic shell \cite{PogorelovLivre}. For small
but not zero $\kappa_i$ values, near the junction, the elastic 
effects smooth out the discontinuity by locally modifying
the shape of the membrane (see Fig.~\ref{fig:coord_fold}) on a 
characteristic distance of order the elastic length $l_e$
in  each phase:
$$l_e = \sqrt{\frac{\kappa_i}{PR_i}}.$$
Using typical values for giant vesicles \cite{Baumgart03}, we get 
$l_e \simeq 0.5 \mu m$, which is very small compared
to $R_i \simeq 10 \mu m$. Therefore, we can model our system as two 
spherical caps slightly distorted at the junction on
a distance of order $l_e$.

\subsubsection{Fold description}

         Far away from the fold, the spherical solution (denoted by S) 
is a good approximation but not in the close
vicinity of the fold better described by a boundary layer (denoted by 
B) of size  $\tilde l_e=l_e/R_i$. We define a new
arclength parameter  $\tilde{l} = 
(\tilde{s}-\tilde{s}_J)/\tilde{l_e}$ and we decompose $\psi$, $r$ and 
$\gamma$ into
\begin{equation}
\label{eqn:fold_decomp_1}
         \psi = \psi_S + \psi_B(\tilde l)\quad\mbox{;}\quad
         r = r_S +\tilde l_e r_B(\tilde l) \quad\mbox{;}\quad
         \gamma = \gamma_S +\tilde l_e \gamma_B(\tilde l).
\end{equation}
with $\psi_S = \theta_1\ \mbox{or}\ \psi_S = \theta_2$. The 
quantities $\psi_B$, $r_B$ and $\gamma_B$ must vanish
far away from the junction. Neglecting absorbed molecules, the 
leading order of Eq.\ref{eqn:ELg} gives
\begin{equation}
\label{eqn:Sol_fold1}
\ddot \psi_B = \sin{\psi_B}.
\end{equation}
This is the pendulum equation with solution:
\begin{equation}
\label{eqn:Sol_fold2}
\tan(\psi_B/4) =\tan(\psi_{cusp}/4)\exp{(\pm \tilde l)}.
\end{equation}
The plus or minus sign is required for $\tilde l$  values, negative 
or positive: after the junction
($\tilde{s} \ge \tilde{s}_J$, $\tilde{l}\ge 0$), or before the 
junction ($\tilde{s} \le \tilde{s}_J$, $\tilde{l} \le 0$).

          From~Eq.\ref{eqn:FTOT} and \ref{eqn:Sol_fold1}, we derive 
the elastic energy in each phase:
\begin{eqnarray}
\label{eqn:Ener_fold1}
         \lefteqn{\frac{F_B}{2\pi PR_i^3} = \tilde{l_e}\sin{\theta_i} 
\left\{ {2\left({1-\cos{\frac{\psi_{cusp}-\theta_i}{2}} } \right) 
}\right. } \\
         && \left. {+ \sin{\theta_i}\left[ { \sin{\theta_i} - 
\sin{\left({\frac{\psi_{cusp}+\theta_i}{2} }\right)} } \right] } 
\right\}. \nonumber
\end{eqnarray}
The elastic energy~(Eq.\ref{eqn:Ener_fold1}), localized at the 
junction is proportional to the interface length ($2\pi 
r_J=2\pi\sin{\theta_i}$ in dimensionless parameters) and has the same 
effect as a line tension. Adding the two contributions, we obtain in 
physical units:
\begin{eqnarray}
\label{eqn:sigma_fold1}
         \lefteqn{\sigma_{cusp}= \sqrt{\kappa_1PR_1} \left\{ 
{2\left[{1-\cos({\frac{\psi_{cusp}-\theta_1}{2}} } )\right] } \right. 
} \\
         && \left. {+ \sin{\theta_1} \left[ { \sin{\theta_1} - 
\sin{\left({\frac{\psi_{cusp}+\theta_1}{2} }\right)} } \right] } 
\right\} + \nonumber \\
         && \sqrt{\kappa_2PR_2} \left\{ {2\left [ 
{1-\cos{\frac{\psi_{cusp}-\theta_2}{2}} } )\right] } \right. 
\nonumber \\
         && \left. {+ \sin{\theta_2} \left[ { \sin{\theta_2} - 
\sin{\left({\frac{\psi_{cusp}+\theta_2}{2} }\right)} } \right] } 
\right\} \nonumber
\end{eqnarray}
The value of $\psi_{cusp}$ is fixed by the boundary 
conditions~(Eq.\ref{eqn:CLg})
\begin{eqnarray}
\label{eqn:psi_cusp}
         \psi_{cusp}=2\arccos\left\{ { 
\frac{\sqrt{R_1\kappa_1}\cos\left({\frac{\theta_1}{2}}\right) + 
\sqrt{R_2\kappa_2}\cos\left({\frac{\theta_2}{2}}\right)}{\sqrt{R_1\kappa_1+R_2\kappa_2 
+2\sqrt{R_1R_2\kappa_1\kappa_2}\cos\left({\frac{\theta_1-\theta_2}{2}}\right)}} 
} \right\}
\end{eqnarray}
As expected $\psi_{cusp}$ depends on the ratio of both rigidities. 
However, it can not be easily measured since
the size of the fold is very small compared to the vesicle size.

          The parameter $\sigma_{cusp}$ measures the strength of 
elasticity on our spherical-cap system. Note that its
contribution
   is angular dependent. Elasticity contributes to the line tension 
and gives an effective line tension
$\tilde{\sigma} = \sigma + \sigma_{cusp}$. However, the total line 
tension is now a function of all the physical
constants ($\sigma$, $P$, $\kappa_o$ and $\kappa_d$) which makes it 
difficult to estimate. Typical values of the elastic
line tensions are $\sigma_{cusp} \simeq 10^{-14} N/m$ (see 
Fig.~\ref{fig:limit_fold1}), for $\sigma \simeq 10^{-13} N/m$.
However, the effect of $\sigma_{cusp}$ on the membrane stability is 
given by the dimensionless number
$$n=\frac{\sigma_{cusp}}{(P\tau_c - \sigma)}$$
which measure the relative effect of the elastic contribution with 
respect to the distance at the bifurcation point.
In our case, we get an important effect with $n=38\%$ and the 
contribution of the elasticity  to the total energy (about
$4\%$) is not enough to affect the zero-order solution, but can be 
important for the fission of the vesicle.

\subsubsection{Effect on the membrane shape.}

         The elastic terms can be taken into account by defining an 
effective line tension.
Therefore, the previous results and the capillary solution are still 
valid but with a new control parameter given by
$\tilde{\tau} = (\sigma + \sigma_{cusp})/P$. Note that the critical 
value $\tau_c$ at the bifurcation is still the same.

\begin{figure}[p]
\begin{center}
          \includegraphics[width=3.25in]{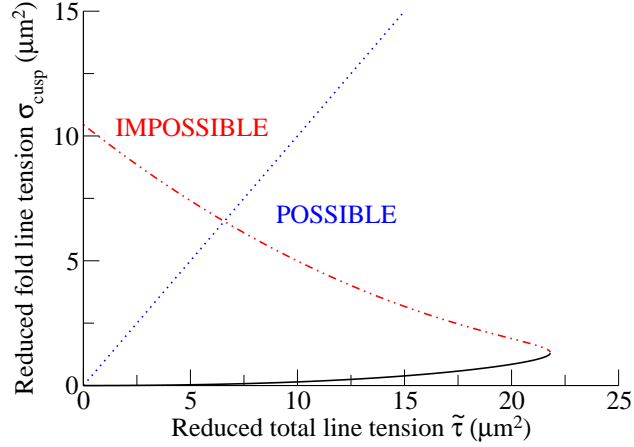}
          \caption{Reduced fold line tension $\sigma_{cusp}$ versus 
the reduced total line tension $\tilde{\tau}$
for the two possible solutions of Eq.\ref{eqn:EL_tense_2}. The solid 
line corresponds to the stable solution, the
dashed line to the unstable solution. The fold line tension has been 
calculated using
Eq.\ref{eqn:sigma_fold1}~and~\ref{eqn:psi_cusp}. The  parameters are 
$A_1 = 136 \mu m^2$, $A_2 = 1296 \mu m^2$ (the
fixed area of each phase), $P = 10^{-2} Pa$, $\kappa_1 = 10^{-19} J$ 
and $\kappa_2 = 10^{-18}J$. The dotted line
separates the possible solutions from the impossible one. For 
$\tilde{\sigma}_{cusp}$ above this line, the line tension
associated to the fold is greater than the total line tension, 
requiring a negative line tension at the junction between
the two domains, which is impossible.
         \label{fig:Fold_tension}}
\end{center}
\end{figure}

\begin{figure}[p]
\begin{center}
          \includegraphics[width=3.25in]{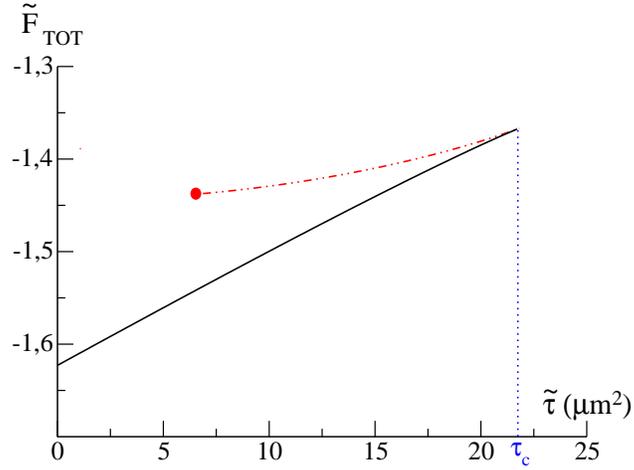}
          \caption{Dimensionless energies $\tilde{F}_{TOT}$ of the 
solutions of the
E.L.~equations~(Eq.\ref{eqn:EL_tense_1} and \ref{eqn:EL_tense_2}) 
versus the control parameter $\tilde{\tau}$ including
the effect of the elastic fold. The parameters are $A_1 = 136 \mu 
m^2$, $A_2 = 1296 \mu m^2$, $P = 10^{-2} Pa$,
$\kappa_1 = 10^{-19} J$ and $\kappa_2 = 10^{-18}J$. The solid line 
corresponds to the stable solution,the dashed line
to the unstable solution, the dotted line to $\tilde{\tau} = \tau_c$, 
the critical value of the
control parameter. For
$\tilde{\tau} \ge \tau_c$, there is no longer a solution.
          \label{fig:ener_fold2}}
\end{center}
\end{figure}

\begin{figure}[p]
\begin{center}
          \includegraphics[width=3.25in]{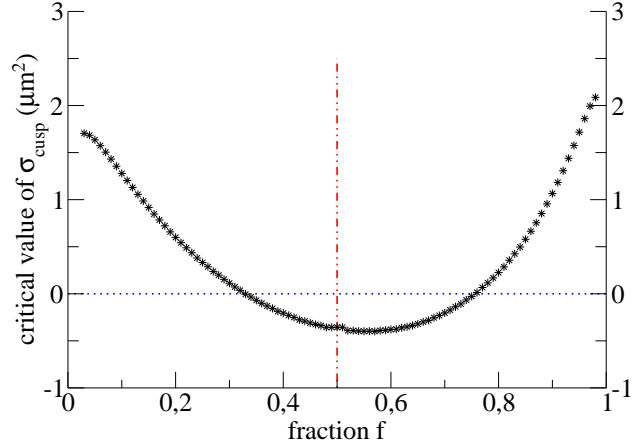}
          \caption{Fold line tension $\sigma_{cusp}$ for the critical 
value of the total line tension
$\tilde{\tau} = \tilde{\tau}_c$, versus the fraction $f$ of the 
liquid-ordered domain. The calculation has been done with
a fixed total area $A_{tot} = A_1 + A_2 = 1433 \mu m^2$, the other 
parameters being $P = 10^{-2} Pa$, $\kappa_1 =
10^{-19} J$ and $\kappa_2 = 10^{-18}J$. The areas of the domains are 
given by $A_1 = f\,A_{tot}$ and $A_2 =
(1-f)A_{tot}$. The dashed line shows the fraction $0.5$ and the 
dotted line $\tilde{\sigma}_{cusp}=0$. The $l_o$
and $l_d$ domains (resp. label $2$ and $1$) do not have the same 
effect since the elastic moduli are not equal. Contrary
to a true line tension, this effective line tension can be negative 
for $f$ near $0.5$.
                \label{fig:limit_fold1}}
\end{center}
\end{figure}

         A variation of the control parameter $\tilde{\tau}$ modifies 
the angles $\theta_1$, $\theta_2$ and
$\psi_{cusp}$ and then the elastic line tension $\sigma_{cusp}$. The 
figure~\ref{fig:Fold_tension} shows the values of
the reduced line tension of the fold ($\tilde{\sigma}_{cusp} = 
\sigma_{cusp}/P$) versus the reduced total line tension
$\tilde{\tau}$. The solid line is $\tilde{\sigma}_{cusp}$ for the low 
energy solution. The dashed line is
$\tilde{\sigma}_{cusp}$ for the high energy solution. The line 
tension of the fold must be smaller than the total line
tension $\sigma + \sigma_{cusp}$ since the line tension $\sigma$ due 
to the interface is positive. Therefore, some shapes
are no longer physically allowed for the unstable solution. The 
figure~\ref{fig:ener_fold2} shows the energies of the
vesicle versus the control parameter $\tilde{\tau}$ for the allowed solutions.

         We have investigated the effect of the size $f$ of the domain 
$1$ on the elastic contribution. In the capillary
model, the two domains are equivalent and $\tau_c$ is the same for 
$f$ and $1-f$. The elasticity breaks this symmetry
since the two domains are no more equivalent: the $l_o$ domain (here 
label $2$) is harder to bend than the $l_d$ one
(here label $1$). The figure~\ref{fig:limit_fold1} shows the elastic 
line tension versus the fraction $f$ of the domain
$1$. Notice that the elastic line tension is negative for large 
domains, meaning that the elasticity fights against
pinching.

\subsection{Budding by molecule insertion}

         The two-cap model remains a solution of the Euler-Lagrange 
equations when molecules are added uniformly.
Eq.\ref{eqn:ELg} connect the concentration of molecules to the 
chemical potential ($\mu_i$) and modify the area Lagrange
multiplier $\Sigma_i$:
\begin{subequations}
\begin{eqnarray}
\mu_i &=& 2\frac{\Lambda_i}{R_i} - \alpha_i(\phi_i-\phi_{eq_i})\\
2\Sigma_iR_i^2 &=&
PR_i^3-2\Lambda_i\phi_iR_i+\alpha(\phi_i^2-{\phi_{eq_i}}^2)R_i^2 \\
\gamma_i &=& \frac{PR_i^2}{2}\sin\psi\cos\psi
\end{eqnarray}
\end{subequations}
In a previous paper, we have shown that the two-cap solutions may be 
unstable either above
a critical homogeneous concentration given by 
$\bar{\phi}_{c_i}=PR_i^2/\Lambda_i$ or for very strong coupling
$\Lambda_i^2/\kappa_i\alpha_i >>1$ \cite{Allain04b}. This instability 
characterizes each phase individually and not the
junction between phases. Here, we focus on the junction and the 
experimental conditions are assumed to be below these
instability thresholds.

\subsubsection{Fold description}

          The interface is the place where strong gradients of 
molecule distribution are found
with typical lengthscale given by
$$l_c =\sqrt{\frac{\beta_i}{\alpha_i}}$$
which must be compared to the vesicle lengthscale $R_i$. We focus on 
the case where $l_c<<R_i$ so
that concentration gradients are also localized at the fold in the 
elastic boundary layer. For distances larger than
$l_c$, the concentration of molecules is constant and reaches the 
value $\bar{\phi}_i$ that we choose as unit in each
phase: so $\tilde{\phi}_i = \phi_i/\bar{\phi}$.

          Far away from the fold, the sphere (denoted by S) is 
solution but not in the vicinity of the fold,
better described by a boundary layer (denoted by B). As previously 
(Eq.\ref{eqn:fold_decomp_1}), we define:
\begin{eqnarray*}
         \tilde{l} &=& 
\frac{(\tilde{s}-\tilde{s}_J)}{\tilde{l}_e}\quad\mbox{;}\quad
         \psi = \psi_S + \psi_B(\tilde l)\quad\mbox{;}\quad
         r = r_S +\tilde l_e r_B(\tilde l) \quad\mbox{;}\\
         \gamma &=& \gamma_S +\tilde l_e \gamma_B(\tilde l)\quad\mbox{and}\quad
         \tilde{\phi} = 1 + \tilde{\phi}_B.
\end{eqnarray*}
To describe the fold, we need  three dimensionless parameters
\begin{equation}
         \tilde{l}_c = \sqrt{\frac{\beta}{\alpha 
R_i^2}}\frac{1}{\tilde{l}_e}, \ \ \
         \tilde{\lambda}_e = 
\frac{\Lambda_i\bar{\phi}_i}{PR_i^2}\frac{1}{\tilde{l}_e}\ \ \mbox{et 
}\
         \tilde{\lambda}_c = \frac{\Lambda_i}{\alpha\bar{\phi}_i 
R_i}\frac{1}{\tilde{l}_e}.
\end{equation}
The conditions for the stability of both phases are 
$\tilde{\lambda}_e \lesssim 1$ and 
$\tilde{\lambda}_e\tilde{\lambda}_c \lesssim 1$. Expanding the shape 
equations~(Eq.\ref{eqn:ELg}) to leading order gives:
\begin{subequations}
\label{eqn:Sol_fold3}
\begin{eqnarray}
\label{eqn:Sol_fold3a}
         \psi_B'' &=& \sin\psi_B + \tilde{\lambda}_e \tilde{\phi}_B', \\
\label{eqn:Sol_fold3b}
         \tilde{l}_c^2\tilde{\phi}_B'' &=& \tilde{\phi}_B - 
\tilde{\lambda}_c \psi_B'.
\end{eqnarray}
\end{subequations}
The fold energy in the phase $i$ is given by the leading orders of 
Eq.\ref{eqn:FTOT}:
\begin{eqnarray}
\label{eqn:Ener_fold_2}
F_i &=& \pi\tilde{l}_e r_J PR_i^3\left[{\int_\mathcal{B}{{\psi'_B}^2d\tilde{l}}
         - \sin\psi_S\int_\mathcal{B}{(\sin\psi}-\sin\psi_S)d\tilde{l}
         - 2\tilde \lambda_e\int_\mathcal{B}\psi'_B d\tilde{l} }\right.
          \nonumber \\
         &&  \left. {+ 4\tilde \lambda_e\int_\mathcal{B}\tilde{\phi}_B
         d\tilde{l} 
-2\tilde{\lambda}_e\int_\mathcal{B}{\tilde{\phi}_B\psi'_B d\tilde{l}} 
+\frac{\tilde{\lambda}_e}{\tilde{\lambda}_c}\int_\mathcal{B}{{\tilde{\phi}_B}^2d\tilde{l}}
+{\tilde{l}_c}^2\frac{\tilde{\lambda}_e}{\tilde{\lambda}_c}\int_\mathcal{B}{{\phi'_B}^2d\tilde{l}} 
}\right].
\end{eqnarray}
The energy is the same for both phases. The sum of the two energies 
is proportional to $r_J$, the interface length,
and defines a new effective line tension $\sigma_{cusp}$.

         The uniform insertion of molecules in the vesicle modifies 
only the Lagrange multipliers which have no direct
physical content, despite the modification of the energy level of the 
system. The two-spherical cap zeroth order solution
remains valid without modification of the geometrical parameters such 
as radii and angles at the junction. Therefore, we
conclude that the bifurcation diagram remains unchanged, except for 
the values of the energy, with the same threshold
value found previously. For $\tau \le\tau_c$, two ideal solutions 
still exist, the stable one being observed
experimentally. Only gradients which appear at the interface modify 
the cusp shapes and we need to evaluate if they are
responsible for a change in the line tension value.

         The equations (\ref{eqn:Sol_fold3}) have no explicit solution 
but some interesting limits can be considered.
  We focus
here on three independent limits: $\tilde{\lambda}_e <<1$, 
$\tilde{\lambda}_c <<1$ and $\tilde{l}_c << \tilde{l}_e$.\\
First case ($\tilde{\lambda}_e <<1$): the elastic coupling length is 
small. This limit decouples Eq.\ref{eqn:Sol_fold3a}
at zero order, giving exactly the same solution as the case without 
molecule. Eq.\ref{eqn:Sol_fold3b} allows to calculate
the molecule distribution but as the terms in $\tilde{\lambda}_e$ can 
be neglected in the energy
(Eq.\ref{eqn:Ener_fold_2}), the effect of the molecules is 
negligible. The elastic line tension is not modified by
molecule addition.\\ Second case ($\tilde{\lambda}_c<<1$): the weak 
chemical coupling length. This limit decouples
Eq.\ref{eqn:Sol_fold3b}, leading to $\tilde{\phi}_B = 
\tilde{\phi}_{B_0}\exp{(\pm\tilde{l}/\tilde{l}_c})$ with
$\tilde{\phi}_{B_0}$ the molecule excess  at the interface, given by 
the boundary conditions~(Eq.\ref{eqn:CLg}). In
physical units and taking into account both sides of the fold, we get 
for the increase of the line tension:
\begin{equation}
\label{eqn:sigma_fold2}
\delta \sigma_{cusp} =
\frac{\sqrt{\alpha_1\beta_1\alpha_2\beta_2}}{2(\sqrt{\alpha_1\beta_1}+\sqrt{\alpha_2\beta_2})}(\bar{\phi}_2-\bar{\phi}_1)^2
\end{equation}
The molecule absorption increases the effective line tension, which 
may induce the fission. This effect is only due
to chemical gradients near the interface. It increases with the 
number of molecules added to the system.\\
Third case ($\tilde{l}_c << 1$): the case of a small chemical length. 
The molecule concentration has two very different
lengthscales: $l_c$ and $l_e$. The chemical length $l_c$ contributes 
to the junction between the two domains and can be
treated as a boundary layer. However, the associated energy is 
proportional to $l_c$ and is then negligible. For size
larger than $l_c$, Eq.\ref{eqn:Sol_fold3} becomes:
\begin{subequations}
\begin{eqnarray}
\lefteqn{\tilde{\phi}_B=\tilde{\lambda}_c\psi'_B}, \\
\lefteqn{(1-\tilde{\lambda}_c\tilde{\lambda}_e)\psi''_B=\sin\psi_B}.
\end{eqnarray}
\end{subequations}
and the effective line tension in the phase $i$ is:
\begin{eqnarray}
\sigma_{cusp} &=& \frac{\sqrt{\kappa_iPR_i^3}}{2}
\left[{(1-\tilde{\lambda}_e\tilde{\lambda}_c)\int_\mathcal{B}{{\psi'_B}^2d\tilde{l}} 
- \sin\psi_S\int_\mathcal{B}{(\sin\psi_B-\sin\psi_S)d\tilde{l}}
}\right. \nonumber \\
&&\left.{ +
4\tilde{\lambda}_e\tilde{\lambda_c}\int_\mathcal{B}{\psi'_Bd\tilde{l}}
- 2\tilde{\lambda}_e\int_\mathcal{B}{\psi'_Bd\tilde{l}}}\right].
\end{eqnarray}
Taking into account the first and second terms leads to the elastic 
line tension~(Eq.\ref{eqn:sigma_fold1})
with a multiplicating factor 
$\sqrt{1-\tilde{\lambda}_e\tilde{\lambda}_c}$ in both phases. Note 
that
$\tilde{\lambda}_e\tilde{\lambda}_c \ge 1$ is not  possible in our 
framework, since the cost associated with the
concentration gradients $\beta$ are neglected \cite{Allain04b}. This 
first contribution indicates that addition of
molecules decreases the line tension associated with the elastic fold 
but since it does not depend on the concentration
it just implies a renormalization of the bending rigidity 
\cite{Leibler86}. The third term, proportional to
$\tilde{\lambda}_e\tilde{\lambda}_c$ also decreases the line tension 
but does not depend on the molecules concentration.
Using physical units, the last integral contributes to the line 
tension by a term proportional to the concentration of
added molecules:
$$\Lambda_1\phi_1(\theta_1-\psi_{cusp}) +
\Lambda_2\phi_2(\psi_{cusp}-\theta_2).$$
It is positive when the molecules are inserted in the outer monolayer 
of the membrane (positive $\Lambda$),
which is the case found in the experiments. However, if the molecules 
are added in the inner monolayer, it becomes
negative and budding and pinching are inhibited. In conclusion, the 
net effect of molecule insertion is a decrease of the
line tension at low concentration, then a possibly increase as the 
concentration increases if the molecules are inserted
from the outer monolayer.

\subsubsection{Budding process}

         The absorption of molecules does not change the zeroth order 
shape equations of the stretched vesicle.
It modifies the shape of the fold near the interface giving a new 
contribution to the effective line tension. If the
absorption takes place in the external leaflet, it contributes to an 
increase of the line tension. This increase puts the
system closer to the bifurcation point controlled by the parameter 
$\tilde \tau=(\sigma + \sigma_{cusp})/P$ and induces a
budding of the smaller phase: as $\tilde{\tau}$ increases, the neck 
radius decreases (see Fig.~\ref{fig:ener_sol1}b) and
the small domain seems to lift up. If the concentration is high 
enough so that $\tilde{\tau} > \tau_c$, the budding is
automatically followed by a fission process, creating two separated 
vesicles, one for each phase. If the concentration is
not high enough, the lift-up will stop before the change of topology. 
In the meantime, it is possible that the vesicle
looses some of its molecules and retracts to its initial 
configuration. This reversibility is impossible when fission is
complete for two reasons: first, the system relaxes the Gaussian 
elastic energy and two daughter vesicles may be
energetically favored, second  due to thermal fluctuations, the 
vesicles move away from each other and the coalescence
process is unlikely. The fact that the fission occurs proves that the 
time scale for fission is much smaller than the
possible rearrangement of molecules between the leaflets

         Figure \ref{fig:staneva1} reproduces experimental results 
from Staneva et al. \cite{Staneva04},
showing fission of a liquid-ordered domain induced by Phospholipase 
${\text A}_{\text 2}$ proteins addition. The vesicle
is obtained by electro-formation (the electrode is visible on the 
left of the pictures). It is composed by a 45:45:10
mol/mol mixture of phosphatidylcholine (PC), sphingomyelin (SM) and 
cholesterol (Chol). The vesicles includes one
liquid-ordered domain visible in fluorescence microscopy (not 
reproduced here): a small fraction ($10\%$) of the PC is
replaced by a fluorescent lipid analog and is excluded from the $l_o$ 
phase, which appears as a dark circle. The proteins
are injected in the neighborhood of the vesicle by a micropipette 
(visible on the right of the first picture).
Phospholipase ${\text A}_{\text 2}$ activity transforms the PC lipids 
into LysoPC, a conical molecule, by cutting one of
the two hydrophobic tails. Fission occurs about 10 seconds after 
protein injection.

         Similar fission process have been observed when detergents 
like LysoPC, Triton X100 or Brij 98 are added in
important quantities near a similar vesicle \cite{Staneva05}. 
However, in this case, the fission is not always complete:
the daughter vesicles may remain connected by a small lipid filament, 
as also observed in \cite{Tanaka04}. This is not in
contradiction with our model since the fission process requires to 
split the lipid bilayers at the molecular level, which
is out of reach of our treatment. This level requires a microscopic 
description as done in \cite{Kozlovsky03}. The
fission process prefers  small domains, as predicted by our model. If 
the concentration in detergents is not high enough
or if the Phospholipase ${\text A}_{\text 2}$ is not activated, the 
liquid-ordered domains bud without complete fission.
It is also possible to observe a relaxation of the vesicle, which 
recovers its initial shape.

\section{Conclusion}

          Our model explains why ejection of a domain from an 
inhomogeneous vesicle can be achieved by osmotic shocks
or molecule absorption. It is based on physical stability concepts in 
the spirit of the existence and stability analysis
of the well-known catenoid.  We predict a complete irreversible 
fission above a critical parameter. From a
macroscopic point of view, the complete fission  is favored, it 
decreases the total energy of the
system at the threshold of stability because of the Gaussian energy. 
This fission can be
inhibited if a membrane thread exists between the two phases. The 
existence of such a thread is out of reach of our
approach. If it does not exist, the vesicles separate from each 
other. If it exists and if the experimental forcing
relaxes, the two vesicles may fuse in principle. The experiments 
discussed here are in favor of a complete fission
mechanism. For simplicity, the model is restricted to two
domains of different sizes: extension to multi-phase domains 
complicates the geometry but will not change the physical
results.

\section*{Acknowledgments}
We would like to thank  M. Angelova, P. Bassereau, T. Baumgart and 
G. Staneva  for various discussions  on
experimental aspects and A. Goriely for critical reading.

\begin{figure}[p]
\begin{center}
          \includegraphics[width=3.25in]{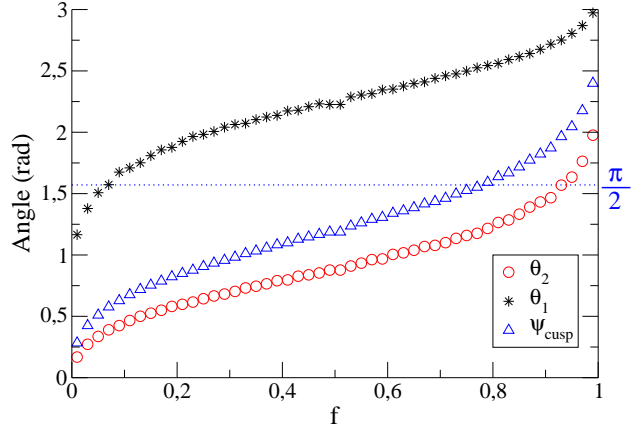}
          \caption{Angles $\theta_1$, $\theta_2$ and $\psi_{cusp}$ for 
$\tilde{\tau} = \tilde{\tau}_c$, the critical value of the control parameter, versus the 
fraction $f$ of the liquid-ordered domain. The calculation
has been done with a fixed total area $A_{tot} = A_1 + A_2 = 1433 \mu 
m^2$, the other parameters being $P = 10^{-2} Pa$,
$\kappa_1 = 10^{-19} J$ and $\kappa_2 = 10^{-18}J$. Label $1$ 
corresponds to the $l_d$ domain and label $2$ to the $l_o$.
The areas of the domains are given by $A_1 = f\,A_{tot}$ and $A_2 = 
(1-f)A_{tot}$. The stars are for the angle
$\theta_1$, the circles for $\theta_2$ and the triangles for $\psi_{cusp}$.
The angle $\psi_{cusp}$ is always closer from the angle $\theta_2$ 
since the liquid-ordered domain is hard to bend.
          \label{fig:psicusp1}}
\end{center}
\end{figure}

\begin{figure}[p]
\begin{center}
          \includegraphics[width=3.25in]{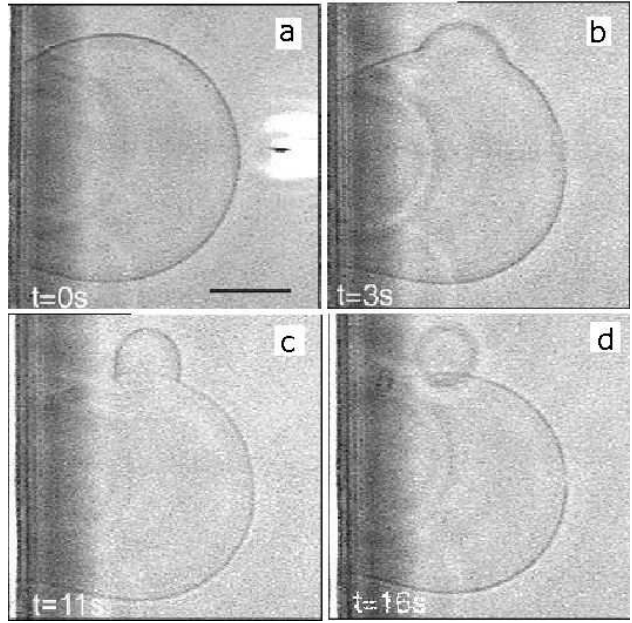}
          \caption{Ejection of a liquid-ordered domain induced by 
proteins Phospholipase A$_2$.
The domain is visible on fluorescence microscopy (not reproduced 
here). The proteins are injected by micropipette (figure
a). The liquid-ordered domain buds (figure b and c) before the 
fission  (figure d). Reproduced form \cite{Staneva04}
with the authorization of the editor. Bar: $20\ \mu m$.
          \label{fig:staneva1}}
\end{center}
\end{figure}


\end{document}